\documentclass[aps,prl,reprint,twocolumn,showpacs,amsmath,amssymb,superscriptaddress]{revtex4}%
\usepackage{multirow}
\usepackage{bigstrut}
\usepackage{amssymb}
\usepackage{amsmath}
\usepackage{graphicx}
\usepackage{dcolumn}
\usepackage{bm}
\usepackage{CJK}
\usepackage{relsize}
\usepackage{cancel}
\usepackage{color}

\begin{document}

\title{Transport Theory of Monolayer Transition-Metal Dichalcogenides through Symmetry}

\author{Yang Song}\email{yang.song@rochester.edu}\affiliation{Department of Physics and Astronomy, University of Rochester, Rochester, New York, 14627}
\author{Hanan Dery}\affiliation{Department of Electrical and Computer Engineering, University of Rochester, Rochester, New York, 14627}\affiliation{Department of Physics and Astronomy, University of Rochester, Rochester, New York, 14627}

\begin{abstract}
We present a theory that elucidates the major momentum and spin relaxation processes for electrons, holes and hot excitons in monolayer transition-metal dichalcogenides. We expand on spin flips induced by flexural phonons and show that the spin relaxation is ultrafast for electrons in free-standing membranes while being mitigated in supported membranes. This behavior due to interaction with flexural phonons is universal in two-dimensional membranes that respect mirror symmetry and it leads to a counterintuitive inverse relation between mobility and spin relaxation.
\end{abstract}
\pacs{}
\maketitle

Single-layer transition-metal dichalcogenides (SL-TMDs) put together exotic charge, spin and valley electronic phenomena in a simple two-dimensional solid-state system \cite{Xiao_PRL12,Cahangirov_PRL12, Lu_PRL13}. Recent advances in characterization of these materials have sparked a wide interest in their $d$-band semiconducting behavior  and spin-valley coupling \cite{Mak_PRL10, Splendiani_Nano10,Korn_APL11,Cao_NatCom12}. Room-temperature mobility of the order of 100~cm$^2$/V$\cdot$s in \textit{n}-type MoS$_2$ monolayer transistor was demonstrated and analyzed \cite{Radisavljevic_NatNano11,Kim_NatCom12,Kaasbjerg_PRB12}. In addition, the unique time-reversal relations between spin and valley degrees of freedom were studied from the exciton photoluminescence (PL) \cite{Zeng_NatNano12,Mak_NatNano12,Sallen_PRB12,Kioseoglou_APL12}. In spite of these recent advances and supporting studies of band-structure and phonon parameters \cite{Fivaz_PR67,Wilson_Yoffe_AP69,Mattheiss_PRB73,Zhu_PRB11,Molina_PRB11,Cheiwchanchamnangij_PRB12}, there is still no unifying description of the spin and charge transport in these materials.

In this Letter, we present a theory that elucidates the intrinsic momentum and spin relaxation mechanisms in SL-TMDs. We first delineate the transport limitations at elevated temperatures via zeroth-order selection rules and make connection with the energy relaxation of hot excitons. It is shown that spin-conserving scattering between direct and indirect exciton bands leads to reduction in the circular polarization degree of the PL. Then, we analyze the intriguing physics of spin relaxation due to scattering with long-wavelength flexural phonons and compare the findings with the case of graphene. Ultrafast spin relaxation of electrons is predicted for free-standing SL-TMDs whereas for supported membranes the spin lifetime is greatly enhanced. Finally, we discuss the relation between spin relaxation and charge mobility.

\begin{figure}[h]
\includegraphics[width=8.5cm]{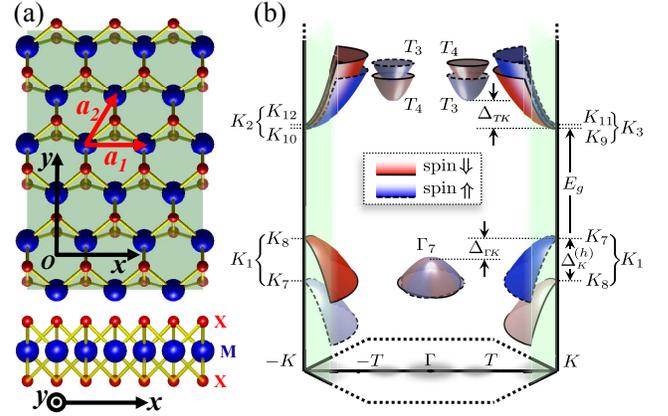}
\caption{(Color online) (a) Top and side views of the real space lattice. (b) Typical scheme of primary and satellite valleys in the conduction and valence bands.} \label{fig:real_recip_lattice}
\end{figure}

\begin{figure}
\includegraphics[width=8cm]{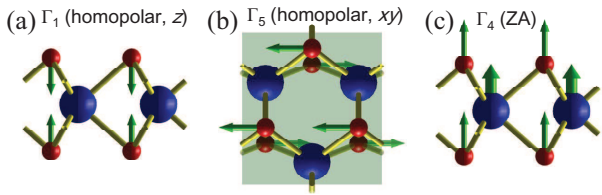}
\caption{(Color online) Atomic displacements of the $\Gamma$~phonon modes involved in zeroth-order scattering.} \label{fig:displacement}
\end{figure}

Figure~\ref{fig:real_recip_lattice}(a) shows the trigonal prismatic lattice structure of monolayer $MX_2$ where $M$ ($X$) denotes the transition-metal (chalcogen) atom. The lack of space-inversion center lowers the symmetry compared with monolayer graphene, leading to spin-split energy bands as shown in Fig.~\ref{fig:real_recip_lattice}(b). Rather than invoking elaborate numerical techniques, we explain the essential transport properties of SL-TMDs by rendering the transformation properties of pertinent electronic states and atomic displacements. This information is captured by the irreducible representations (IRs), shown in Fig.~\ref{fig:real_recip_lattice}(b) for edge states of the conduction and valence bands \cite{supple}. We begin with the electron-phonon interaction in its zeroth order, characterized by nonvanishing scattering amplitude between band-edge states \cite{Song_PRB12}. In  experiments, these processes are identified by their temperature dependence (Bose-Einstein distribution of the involved phonon). Table~\ref{tab:zeroth_order_scattering} lists the selection rules, and Fig.~\ref{fig:displacement} shows the underlying atomic displacements for cases of zone-center phonons. With the exception of $\Gamma_4$ that corresponds to zeroth-order spin flips in the $T$~valleys, energies of all other phonon modes are non-negligible \cite{supple}. Thus, only at elevated temperatures the transport in the $K$~valleys is affected by zeroth-order processes.

\begin{table}
\renewcommand{\arraystretch}{1.4}
\tabcolsep=0.2cm
\caption{\label{tab:zeroth_order_scattering} Zeroth-order selection rules in MX$_2$ compounds. For spin-conserving scattering, double-group IRs are replaced with simpler single-group IRs. Time reversal symmetry connects $K$ and $-K$~points (e.g., $K_{9}$=$K_{10}^{\ast}$ and $K_{2}$=$K_{3}^{\ast}$). } 
\begin{tabular}{l|ll}
\hline \hline
Valleys                                     & \quad spin-conserving                            &  $\qquad$ spin-flip \\ \hline
\multirow{2}{*}{Intra}                & $(X\times X^*)^*=\Gamma_1$                 &  $(K_{11} \times K^*_9)^* = \Gamma_5$ \\
                                            & $X\!=\!\{K_{1-3}, T_1, \Gamma_1 \}$           &  $(T_3 \times T^*_4)^*=\Gamma_4$   \\ \hline
\multirow{4}{*}{Inter}                &  $(K_3\times K^*_2)^* =K_3$                &                                  \\
                                            &  $(K_1\times K^*_1)^* =K_1$                &                                  \\
                                            &  $(T_1\times K^*_3)^* =T_1$                &  $(T_3\times K^*_{9})^*=T_2$    \\
                                            &  $({\Gamma_1 \times K^*_1 })^* = K_1$      &  $( \Gamma_{7\downarrow}\times K^*_{7})^*=K_{6}$ \\ \hline \hline
\end{tabular}
\end{table}

We summarize the physics encompassed in Table~\ref{tab:zeroth_order_scattering}. The intravalley relaxation of either electrons or holes is induced by homopolar scattering due to an out-of-phase displacement of the two chalcogen atomic layers. As implied from Figs.~\ref{fig:displacement}(a) and (b) for the respective cases of momentum and spin relaxation, the relaxation is governed by short-range scattering (i.e., the effective charge dipole from these long-wavelength displacements is zero). The intravalley momentum relaxation is caused by thickness fluctuations of the layer due to the out-of-plane motion of the chalcogen atoms [Fig.~\ref{fig:displacement}(a)]. This physical picture was first identified by Fivaz and Mooser \cite{Fivaz_PR67}, and supported by \textit{ab initio} calculations of Kaasbjerg \textit{et al}., who also showed a comparable contribution to the charge mobility from Fr\"{o}hlich interaction \cite{Kaasbjerg_PRB12}. Zeroth-order spin flips in the $K$~valleys are enabled uniquely by homopolar in-plane optical phonons which do not exist in graphene structures [Fig.~\ref{fig:displacement}(b)].

The intervalley scattering between primary and satellite valleys ($K$$\times$$T$ \& $K$$\times$$\Gamma$) is relevant due to the flat nature of the $d$~bands in SL-TMDs. This scattering is likely to facilitate the Gunn effect when applying a large in-plane electric field \cite{Gunn_IBM64}. Namely, accelerated carriers are scattered to the satellite valleys in which the mobility and spin relaxation rates are different \cite{Kroemer_IEEE64, Li_PRL12}. $MX_2$ compounds with heavy (light) chalcogen atoms have a relatively small $\Delta_{T K}$ ($\Delta_{\Gamma K}$) energy spacing \cite{Zhu_PRB11}, and therefore can be used as $\textit{n}$-type ($\textit{p}$-type) Gunn diodes. Finally, we discuss the nonvanishing intervalley scattering between edges of the $K$ and $-K$ valleys.  The selection rules show that such scattering largely affects the charge but not the spin transport. The lowest-order spin-flip of either electrons or holes is forbidden by time reversal symmetry. The conduction-band rule ($K_3$$\times$$K^*_2$) is relevant for electron transport in $n$-type monolayers \cite{Kaasbjerg_PRB12}. The valence-band rule ($K_1$$\times$$K^*_1$), however, is likely to affect less the transport of holes due to the relatively large energy splitting ($K_8$$\,$$\leftrightarrow$$\,$$K_8$ or $K_7$$\,$$\leftrightarrow$$\,$$K_7$ intervalley transitions in Fig. \ref{fig:real_recip_lattice}).


\begin{figure}
\includegraphics[width=8.5cm]{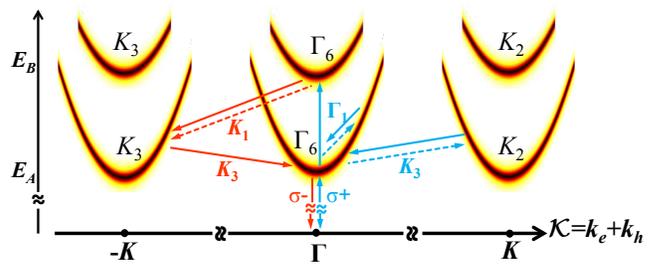}
\caption{(Color online) Direct and indirect exciton bands.  The optical and exciton-phonon scattering processes are sketched under $\sigma_+$ photon excitation. Real (virtual) scattering is denoted by solid (dashed) lines and phonon IRs are marked. Right blue (left red) paths lead to $\sigma_+ (\sigma_-)$ luminescence.} \label{fig:exciton}
\end{figure}

As an important application, we show that spin-conserving scattering between $K$ and $-K$ valleys is imperative for understanding recent exciton PL measurements \cite{Zeng_NatNano12,Mak_NatNano12,Sallen_PRB12,Kioseoglou_APL12}. Figure~\ref{fig:exciton} shows the exciton bands of singlet composites (``bright excitons''), where upper and lower branches are due to the valence-band spin splitting \cite{footnote_screening}. The zone-center bands comprise electron and hole states of the same $K$ valley (direct excitons), where each of the doubly-degenerate zone-center states transform as $\Gamma_6$ and include $m_l$=$\pm$1 excitons depending on light helicity. Zone-edge bands, on the other hand, belong to $K_{2,3}$ and comprise electrons and holes from opposite $K$ valleys (indirect excitons). In the supplemental material we quantify the circular polarization degree of the PL by modeling the absorption and relaxation processes that precede recombination. Here, we summarize this physics and show the relation with selection rules. First, when the exciting photon energy is between $E_A$ and $E_B$ (Fig.~\ref{fig:exciton}), both upper and lower bands are excited due to energy broadening by impurities and substrate imperfections \cite{Kioseoglou_APL12,Zeng_NatNano12,Mak_NatNano12}. Second, the phonon-assisted indirect absorption, a second-order process shown by dashed arrows in Fig.~\ref{fig:exciton}, cannot be neglected since it has many more available final states compared with the direct absorption that is limited to $\bm{\mathcal{K}}$$\,$=$\,$$\mathbf{k}_e$+$\mathbf{k}_h$$\,$=$\,$0 \cite{Elliott_PR57}. Third, the intervalley scattering during the relaxation from the upper to lower $\Gamma_6$ band flips $m_l$ without a spin-flip of the electron or hole \cite{footnote_unique}. The missing angular momentum is carried by $K_1$ and $K_3$ phonons, as implied from Table~\ref{tab:zeroth_order_scattering} and visualized in Fig.~\ref{fig:exciton} for $\sigma_+$ excitation. This rapid relaxation explains the measured reduction in the circular polarization degree \cite{Kioseoglou_APL12}, and it is enabled in multivalley direct-gap SL-TMDs due to the (so-far ignored) unique coexistence of direct and indirect exciton bands.

In the remainder of the Letter, we study the transport due to interaction with long-wavelength acoustic phonons. As in other materials, the vanishing energies of these phonons render this nonzero-order interaction important. We first mention the fundamental distinction between charge and spin transport in monolayers that respect mirror symmetry (e.g., graphene and SL-TMDs). For interactions with single phonons, spin-conserving scattering is not affected by long-wavelength flexures in the out-of-plane direction \cite{Mariani_PRL08}, whereas spin-flip scattering induced by flexural phonon does not vanish [the ZA mode in Fig.~\ref{fig:displacement}(c)]. Mathematically, it is understood from the scattering integral, $\langle \mathbf{s}_f| \nabla \mathcal{V} | \mathbf{s}_i \rangle$,  where in-plane $\partial \mathcal{V} / \partial r_\|$ (out-of-plane $\partial \mathcal{V} / \partial z$) deformations are even (odd) with respect to mirror symmetry which also brings in $\pm 1$ for $\mathbf{s}_i\!= \!\pm\mathbf{s}_f$.


We focus on the Elliott-Yafet spin-flip mechanism due to the intrinsic electron-phonon scattering \cite{EY_mechanism,footnote_mom}. Using the method of invariants \cite{Bir_Pikus_Book}, the spin-flip transition amplitudes from scattering with flexural phonons are compactly expressed as
\begin{eqnarray}
\mathcal{T}_{\scriptscriptstyle{ZA},\mathbf{q}}^{sf} \!\approx\! \sqrt{\frac{k_BT}{2\varrho A \gamma^2q^{2\eta} }}\,g(\mathbf{k}_i,\!\mathbf{k}_f)\,, \label{eq:Tsf}
\end{eqnarray}
where $\varrho$ is the areal mass density and $A$ is the area. $\gamma$ and $\eta$ are mechanically dependent parameters that set the flexural-phonon dispersion. Their values will be introduced when estimating the spin lifetimes in different monolayer conditions. $\mathbf{q}=\mathbf{k}_i-\mathbf{k}_f$ denotes the small phonon wave vector where $\mathbf{k}_i$ and $\mathbf{k}_f$ are wavev ectors of the initial and final electronic state, respectively. $g(\mathbf{k}_i,\!\mathbf{k}_f)$ depends on the valley position and reads $\Xi_{\Gamma_1}^{so}$$|\mathbf{k}_i+\mathbf{k}_f|q$ in the zone center, $\Xi_{K}^{so}$$q$ in the zone edge, or $\Xi_{T_1}^{so}$ inbetween. $\Xi_{X}^{so}$ are spin-orbit coupling scattering constants \cite{supple}. The linear wave vector dependence of $g(\mathbf{k}_i,\!\mathbf{k}_f)$ in the zone edge enables estimation of $\Xi_{K}^{so}$ from the energy change of the $K$-point spin splitting in response to static strain. By symmetry, shear-strain components of the form $\epsilon_{\pm}$$\,$=$\,$$\epsilon_{xz}\!\pm\! i\epsilon_{yz}$ are associated with frozen flexural phonons. By focusing on the spin dependent part of the static strain Hamiltonian at $\mathbf{k}=\mathbf{K}$,
\begin{eqnarray}
H_0+ H^{so}(\epsilon_+)= \left(\begin{array}{cc}  b |\epsilon_+|^2 & a\epsilon_+ \\ a \epsilon_- & \Delta_K - b |\epsilon_+|^2 \end{array}\right), \label{eq:strain}
\end{eqnarray}
we find that $a$$\,$=$\,$$\Xi^{so}_K$ \cite{supple}. $a$ and $b$ are spin-dependent shear deformation potentials, and $\Delta_K$ is the spin splitting without strain. The value of $\Xi^{so}_K$, needed for estimation of the spin relaxation, can now be readily extracted from the strain-induced spin splitting $\sqrt{(\Delta_K+2 b|\epsilon_+|^2)^2+ |2a\epsilon_+|^2}$. Using ABINIT (an open-source DFT code) with Hartwigsen-Goedecker-Hutter pseudopotentials, we find that $\Xi^{so}_K$ for monolayer MoS$_2$, MoSe$_2$, WS$_2$ and WSe$_2$ are, respectively 0.2, 0.27, 0.66 and 0.67~eV in the conduction band \cite{supple, Xi_so_value}, and similar magnitudes in the valence band. A notable feature in the conduction band of SL-TMDs is that $\Delta_K$ can be much smaller than $\,$$\Xi^{so}_K$: whereas $\Delta_K$ is mostly governed by the small deviation of the state from centrosymmetric $d_{z^2}$-like orbitals (for which the spin-splitting vanishes),  $\Xi^{so}_K$ stems from \textit{interband} spin-orbit coupling between different $d$ orbitals \cite{supple}.

Having values of the spin-dependent scattering constants, we quantify the $K$-valley spin relaxation rate due to interaction with flexural phonons. For electrons, its contribution dominates all other processes  in Table~\ref{tab:zeroth_order_scattering}.  Using Fermi golden rule with (\ref{eq:Tsf}) and assuming elastic scattering, the spin-flip rate of the $\mathbf{k}$ state is
\begin{eqnarray}
\tau_{s}^{-1}(\mathbf{k}) \approx \frac{2m_{t(\ell)}(\Xi^{so}_{\scriptscriptstyle{K}})^2k_BT }{\hbar^3\varrho\gamma^2 (k' \!\!+\! k)^{{2\eta-2}}} \cdot{}_2F_1\!\left(\!\tfrac{1}{2}, \eta\!-\!1; 1; \tfrac{4kk'}{(k + k')^2}\!\right)\!. \label{eq:Tsf_of_k}
\end{eqnarray}
$\ell$$(t)$ denotes scattering from the top to lower spin-split bands (or vice versa). $m_{t(\ell)}$ is the effective mass and $k'=\sqrt{m_{t(\ell)} k^2/m_{\ell(t)}+(-) 2m_{t(\ell)} \Delta_{\scriptscriptstyle{K}}/\hbar^2}$. The hypergeometric function can be recast to simpler forms for case-specific $\eta$ values. The respective expression for $T$~valley spin flips is similar in form to (\ref{eq:Tsf_of_k}), but with $\eta$$\,$$\rightarrow$$\,$$\eta$+1 which reflects faster spin relaxation (as implied from Table~\ref{tab:zeroth_order_scattering}). We continue the analysis and calculate the $K$-valley spin relaxation rate in two limiting cases.

\textit{Free-standing monolayers}.---Without a stiffening mechanism to suppress violent undulations, two-dimensional membranes would crumple \cite{Nelson_JPP87}. In crystal monolayers, such mechanism is naturally provided by the coupling between bending and stretching degrees of freedom. The coupling, in the lowest order that satisfies flat phase conditions, renormalizes the dispersion power law of long-wavelength flexural phonons from $\eta$$\,$=$\,$2 to $\eta$$\,$=$\,$3/2, and it yields $\gamma$$\,$=$\,$$\sqrt[4]{k_BT/\varrho v_0^2}$ where $v_0 \approx v_{\scriptscriptstyle{TA}} \sqrt{1-v_{\scriptscriptstyle{TA}}^2 / v_{\scriptscriptstyle{LA}}^2}$ is expressed in terms of the mode-dependent sound velocities \cite{Nelson_JPP87,Mariani_PRL08}.  Using these parameters and assuming $m_{\scriptscriptstyle{K}}$$\,$$\equiv$$\,$$m_t$$\,$=$\,$$m_{\ell}$, the effective spin relaxation rate for Boltzmann distribution in the spin-split bands becomes
\begin{eqnarray}
\frac{1}{\tau_{s}}= \sqrt{\frac{8\pi m_{\scriptscriptstyle{K}} }{\varrho v_o^2}} \left( \frac{\Xi^{so}_{\scriptscriptstyle{K}}}{\hbar}\right)^{\!2}\left[\frac{\rm{Erfc}(\sqrt{\beta_K})}{1 + e^{-\beta_{\scriptscriptstyle{K}}}} \left( 1 + \frac{0.1}{\sqrt{\beta_{\scriptscriptstyle{K}}}}\!\right) \right],\,\,\,\,\label{eq:tau_sf}
\end{eqnarray}
where $\beta_K$$\,$=$\,$$\Delta_{\scriptscriptstyle{K}}$/$k_B$$T$. Due to a relatively large valence-band splitting, the $\beta_K$$\,$$>$$\,$1 limit applies for holes at all practical cases where the temperature dependence is largely set by the complementary error function. Furthermore, in compounds with heavier transition-metal atoms (larger splitting), the flexural induced spin relaxation of holes is slower in spite of a larger $\Xi^{so}_{\scriptscriptstyle{K}}$. For example, in MoS$_2$ where $\Delta_{\scriptscriptstyle{K}}^{(h)}$$\,$=$\,$160~meV and in WS$_2$ where $\Delta_{\scriptscriptstyle{K}}^{(h)}$$\,$=$\,$450~meV \cite{Zhu_PRB11}, the flexural induced spin lifetimes at 300~K are, respectively, $\sim$$\,$0.04~ns and $\sim$$\,$0.5~$\mu$s \cite{footnote_params}. In addition to scattering with flexural phonons, the intrinsic spin relaxation of holes is affected by intravalley scattering with in-plane homopolar phonons [Fig.~\ref{fig:displacement}(b)], or by intervalley scattering between $K$ and $-K$. Whereas the latter spin-flip scattering is forbidden in the zeroth order by time reversal symmetry (Table~\ref{tab:zeroth_order_scattering}), it is not impeded by the relatively large spin splitting. Signatures of the homopolar and intervalley spin-flip mechanisms can be observed from their temperature dependence.

The spin relaxation of electrons is much faster due to the small spin splitting in the conduction band.  In MoS$_2$ where $\Delta^{\!(e)}_K \approx 4$ meV \cite{Cheiwchanchamnangij_PRB12}, the room-temperature spin lifetime  is $\tau_{s}$$\,$$\sim$$\,$0.05~ps \cite{footnote_params}, and it increases noticeably only below 50~K. Interestingly, the spin relaxation of  electrons is enhanced in compounds with lighter metal atoms [their weaker spin-orbit coupling leads to smaller $\Delta^{(e)}_K$]. Furthermore, the spin relaxation rate diverges in the pathological limit $\Delta^{\!(e)}_K\rightarrow0$ \cite{footnote_mobility}.

\begin{figure}
\includegraphics[width=8.5cm]{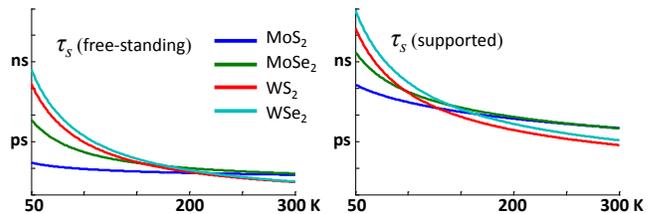}
\caption{(Color online) Electron spin lifetimes versus temperature, governed by intravalley scattering with flexural phonon for free-standing [Eq.~(\ref{eq:tau_sf})] and supported SL-TMDs [Eq.~(\ref{eq:tau_sf_sup})].} \label{fig:tau_s}
\end{figure}

\textit{Supported monolayers}. ---Another means to stiffen the membrane is naturally provided by van der Waals (vdW) interactions when the monolayer is placed on a substrate \cite{Seol_Science10}. The support brings in a minimum cutoff energy for out-of-plane displacements. The cutoff energy, $\Omega_c$ $\,$=$\,$$\hbar$$\,$$\sqrt{\kappa_s/\scriptstyle{M}_u}$, is calculated from the average vdW interatomic force constant between the monolayer and the substrate ($\kappa_s$), and the average atomic mass of the monolayer ($\scriptstyle{M}_u$). In the long-wavelength limit, we can therefore approximate the dispersion of flexural phonons by $\eta$$\,$=$\,$0 and $\gamma$$\,$=$\,$$\Omega_c$/$\hbar$. The effective spin relaxation rate becomes
\begin{eqnarray}
\frac{1}{\tau_{s}}= \frac{ \left(2 m_{\scriptscriptstyle{K}} \Xi^{so}_{\scriptscriptstyle{K}}\right)^2 }{\hbar^3\varrho} \cdot \frac{4+2\beta_K}{(1 + e^{\beta_{\scriptscriptstyle{K}}})\beta_c^{2}} \,,\,\,\,\label{eq:tau_sf_sup}
\end{eqnarray}
where $\beta_c$$\,$=$\,$$\Omega_c$/$k_B$$T$$\,$$<$$\,$1 is assumed. The temperature dependence is quadratic for $\beta_c$$\,$$<$$\,$$\beta_K$$\,$$<$$\,$1 and exponential for $\beta_c$$\,$$<$$\,$1$\,$$<$$\,$$\beta_K$. The substrate coupling brings in slower relaxation that at room temperature reaches $\tau_{s}$$\,$$\sim$$\,$3~ps ($\sim$$\,$0.2~ns) for electrons (holes) in supported MoS$_2$ with $\Omega_c\approx$1~meV \cite{footnote_force_constant}. The spin-lifetime enhancement from $\eta$$\,$=$\,$3/2 to $\eta$$\,$=$\,$0 is sharper for electrons due to their smaller spin-splitting (phonons with longer wavelength are capable of inducing transitions between opposite spin bands). Figure~\ref{fig:tau_s} summarizes the temperature dependence of $\tau_s$ for various SL-TMDs \cite{footnote_params,footnote_force_constant}.

We compare the spin relaxation induced by flexural phonons in graphene and MX$_2$. The space inversion symmetry in  graphene mandates spin-degenerate energy bands, resulting in anisotropic relaxation that depends on the spin orientation. Defining the latter by out-of-plane polar angle ($\theta_{\mathbf{s}}$) and in-plane azimuthal angle ($\phi_{\mathbf{s}}$), the spin-flip transition amplitude due to elastic scattering with flexural phonons follows (\ref{eq:Tsf}) where
\begin{eqnarray}
\!g(\mathbf{k}_i,\!\mathbf{k}_f) \!= \! i k \Xi_{\mathbf{\scriptscriptstyle{K}}}^{so} \sin^{2}\!\!\phi_{\!-} \!\!\left[ \sin^{2}\!\frac{\theta_{\mathbf{s}}}{2}e^{2i\phi_{\mathbf{s}}- i\phi_{\!{+}}}\!+\cos^{{2}}\!\frac{\theta_{\mathbf{s}}}{2}e^{i\phi_{\!{+}}}\right]\!.\,\label{eq:Tsf_gr}
\end{eqnarray}
The other angles, $\phi_\pm$$\,$=$\,$$\tfrac{1}{2}$($\tan^{-1}{\tfrac{k_{x,f}}{k_{y,f}}}$$\,$$\pm$$\,$$\tan^{-1}{\tfrac{k_{x,i}}{k_{y,i}}}$), are due to the Dirac-cone energy dispersion. Formal analytical derivation of this result will be presented in a future long publication. Here we mention that the prefactor $k\sin^2\phi_-$=$\,q^2/4k$, previously found by numerical techniques \cite{Fratini_arXiv2012}, originates from space-inversion symmetry and spin-dependent energy dispersion in the cone region. Away from the Dirac point ($k$=0), the spin relaxation times cales are longer in graphene than in SL-TMDs on accounts of the higher power law in phonon wave vector ($q^2$ vs $q$) and the relative smallness of $\Xi_{\mathbf{\scriptscriptstyle{K}}}^{so}$ in carbon-based systems ($\sim$$\,$10~meV \cite{Hernanso_PRB06}). 

All these findings lead to a counterintuitive relation between charge mobility and spin relaxation in two-dimensional membranes. Whereas increased stiffness has been shown to be associated with slower spin relaxation, its coupling with charge mobility seems to have the opposite trend. For example, high mobility in supported membranes is a token of diminished effect from adsorbents and substrate imperfections. A smaller coupling of the membrane to such parasitics would enable freer and softer out-of-plane undulations leading to ultrafast spin relaxation of electrons without affecting their mobility (forward spin-flip scattering by long-wavelength flexural phonons).  Therefore, an inverse trend between spin and momentum relaxation, a hallmark of Dyakonov-Perel spin dephasing processes, can be realized in a Elliott-Yafet spin flip system. This physics is universal in relatively clean two-dimensional membranes that respect mirror symmetry. In such membranes, charge transport is decoupled from harmonic out-of-plane undulations while spin relaxation is not severely affected by the presence of impurities.



In conclusion, we have presented a concise theory of intrinsic transport properties in single-layer transition-metal dichalcogenides. Lowest-order scattering processes were identified from group theory in both charge and spin transport regimes, and were found relevant to the transport of free-carriers at elevated temperatures and for the energy relaxation of hot excitons. In addition, the spin relaxation induced by scattering with long-wavelength flexural phonons was quantified. For electrons, the ultrafast rate is attributed to the typical softness of two-dimensional membranes and to the small spin splitting in the conduction band. The relatively large splitting in the valence band, on the other hand, renders $p$-type monolayers better candidates for preserving spin information. 

We are indebted to Professor Walter Lambrecht and Mr. Tawinan Cheiwchanchamnangij for insightful discussions and for sharing frozen-phonon DFT results prior to their publication. This work is supported by NRI-NSF, NSF, and DTRA Contracts N. DMR-1124601, No. ECCS-1231570, and No. HDTRA1-13-1-0013, respectively.


\begin{thebibliography}{99}
\bibitem{footnote_screening} The reduced screening in two-dimensional renders a relatively large exciton binding energy such that $E_{A,B}$ in Fig.~\ref{fig:exciton} lie well below the one-particle energy gaps \cite{Cheiwchanchamnangij_PRB12,Mak_NatNano12}.
\bibitem{footnote_mobility} By repeating the analysis for suspended monolayers for which $\eta$$\,$=$\,$1 and $\gamma=v_0\sqrt{\overline{\epsilon}}$, we find that  $\tau_{s}$$\,$$<$$\,$1~ps at 300~K even for suspension-induced strain levels of $\overline{\epsilon}\sim 1\%$.
\bibitem{footnote_force_constant} We assumed a force constant of $\sim$$\,$0.4~N/m, similar to the one found in graphene on SiO$_2$ \cite{Seol_Science10}. We use similar cutoff energy $\Omega_c\approx$1 meV for other compounds in Fig.~\ref{fig:tau_s}.
\end{thebibliography}
\end{document}


\title{Supplemental material for ``Transport Theory of Monolayer Transition-Metal Dichalcogenides through Symmetry''}

\author{Yang Song ËÎÑï}\affiliation{Department of Physics and Astronomy,
University of Rochester, Rochester, NY 14627}
\author{Hanan Dery}\affiliation{Department of Electrical and Computer Engineering, University of Rochester, Rochester, NY 14627}\affiliation{Department of Physics and Astronomy,
University of Rochester, Rochester, NY 14627}

\maketitle
{\bf 1. Determination of electron band representations in Fig.~1, identification of phonon modes in Table~1 and Fig.~2, and character tables.} A general periodic Bloch wavefunction can be written as
\begin{eqnarray}\label{eq:tight_binding_state}
\Phi_{\mathbf k, n}(\mathbf r) = \sum\limits_j e^{i\mathbf k\cdot \mathbf R_{j,\alpha}} \phi_{\mathbf k n}(\mathbf r-\mathbf R_{j,\alpha}),
\end{eqnarray}
where $\mathbf k$ and $n$ index the wavevector and band in the reduced Brillouin zone. $R_{j,\alpha}$ is the lattice site at the $j$th primitive cell and $\alpha$ atom, and $\phi_{\mathbf{k}n}$ is the associated atomic wavefunction. The symmetry of $\Phi_{\mathbf k, n}$ is thus determined by both the exponential factor and $\phi_{\mathbf{k}n}$, and complies with the irreducible representation (IR) of the given wavevector group.  For definiteness, the coordination in Fig.~1 of the main text is used and the operations are defined such that $C^+_3$ rotates the wavefunction $2\pi/3$ counterclockwise (or the coordinate $2\pi/3$ clockwise). Physical results are independent of such choices.

We demonstrate in some detail the assignment of conduction and valence band state IRs at $\mathbf{K}$ point, by comparing its character table (Table~\ref{tab:CharactorTable_K}) and Eq.~(\ref{eq:tight_binding_state}). Results at other high symmetry points in Fig.~1, as well as the IRs for phonon polarization vectors, can be done using this method (with Tables~\ref{tab:CharactorTable_Gamma}-\ref{tab:CharactorTable_T}).  We start with the well-known facts that the transition-metal orbitals at $\mathbf{K}$ point are mostly $d_{z^2}$ for  the conduction band and $d_{xy}$ and $d_{x^2-y^2}$ for the valence band \cite{Wilson_Yoffe_AP69}. Spin-independent IR $K_3$ for conduction band immediately follows, noting that the exponential factor in Eq.~(\ref{eq:tight_binding_state}) is determined by the lattice sites. The ambiguity of the relative phase in the valence band can be resolved by simply noting the fact, that both valence spin-split bands can have optical transition with conduction bands (two bright excitons). This fact forces the IRs to be $K_{9(11)}$ for spin $\uparrow_z(\downarrow_z)$, with the fixed phase in orbital, $d_{x^2-y^2}+i d_{xy}$. Here we point out that all of the IRs of the group at $\mathbf{K}$ are one-dimensional, and thus suggest \textit{no spin-degenerate bands}.

The energy orders of the spin-split bands may be determined by the orbital magnetic numbers. For example at $\mathbf{K}$ point valence band, $m_l=2$ for $d_{x^2-y^2}+i d_{xy}$ determines that  $m_s=\frac{1}{2}$ occupies higher energy band due to positive spin-orbit coupling coefficient. The spin order in the conduction bands is not so clear. Two small contributions to the spin splitting compete: around the chalcogen core, the small $p_x-i p_y$ (required by $K_3$ IR) component with $m_l=-1$ raises $m_s=-\frac{1}{2}$ band; around the metal core, the wavefunction deviates from inversion symmetric shape (as $d_{z^2}$ or $s$) due to the triangular lattice potential, and this deviation can induce spin-orbit coupling. DFT calculation indicates that the influence from the metal core dominates in conduction band spin-orbit splitting (and $m_s=\frac{1}{2}$ occupies the higher conduction band) except for the MoSe$_2$ case. Caution should be used when connecting $K$ group IRs with magnetic numbers, which have only three-fold symmetry.

\begin{table}[!htbp]
{ 
\caption{Character table at $\Gamma$ point, the $D_{3h}$ point double group \cite{Bradley_Cracknell72}. }
\label{tab:CharactorTable_Gamma}
\renewcommand{\arraystretch}{1.1}
\tabcolsep=0.03cm
\begin{tabular}{cc|ccc ccc ccc}
\hline \hline
\multicolumn{2}{c|}{$D_{3h}$} &$E$& $\bar{E}$& $C^+_3\,C^-_3$& $\bar{C}^+_3\,\bar{C}^-_3$  &$\sigma_h\,\bar{\sigma}_h$ &$S^+_3\,S^-_3$ & $\bar{S}^+_3\,\bar{S}^-_3$ & $C'_{2i} \bar{C}'_{2i}$ & $\sigma_{vi} \bar{\sigma}_{vi}$  \\ \hline
$A'_1$ &$\Gamma_1$            & 1 & 1        &1              &1                            & 1                          & 1            &1                           &1                        &  1
  \\
$A'_2$ &$\Gamma_2$            & 1 & 1        &1              &1                            & 1                          & 1            &1                           &$-1$    &$-1$
  \\
$A''_1$ &$\Gamma_3$           & 1 & 1        &1              &1                            &$-1$                        &$-1$          &$-1$                         &1                        &$-1$     \\
$A''_2$ &$\Gamma_4$           & 1 & 1        &1              &1                            &$-1$                        &$-1$          &$-1$                         &$-1$                        &$1$
  \\
$E''$ &$\Gamma_5$             & 2 & 2        &$-1$           &$-1$                         &$-2$                        &1             &1                            &0                        &0
\\
$E'$ &$\Gamma_6$              & 2 & 2        &$-1$           &$-1$                         &$2$                        &$-1$            &$-1$                        &0                        &0
\\
$\bar{E}_1$ &$\Gamma_7$       & 2 &$-2$        &$1$         &$-1$                         &0                        &$\sqrt{3}$         &$-\sqrt{3}$                 &0                        &0
 \\
$\bar{E}_2$ &$\Gamma_8$       & 2 &$-2$        &$1$         &$-1$                         &0                        &$-\sqrt{3}$         &$\sqrt{3}$                 &0                        &0
    \\
$\bar{E}_3$ &$\Gamma_9$       & 2 &$-2$        &$-2$         &$2$                         &0                        &0                  &$0$                         &0                        &0     \\
\hline\hline
\end{tabular} }
\end{table}

\begin{table}[!htbp]
{ 
\caption{Character table at $K$ point, the $C_{3h}$ point double group \cite{Bradley_Cracknell72}. We replace $\Gamma$ with $K$  in the IR notation.}
\label{tab:CharactorTable_K}
\renewcommand{\arraystretch}{1.2}
\tabcolsep=0.03cm
\begin{tabular}{cc|ccc ccc ccc ccc}
\hline \hline
\multicolumn{2}{c|}{$C_{3h}$} &$E$& $C^+_3$& $C^-_3$ &$\sigma_h$ &$S^+_3$ & $S^-_3$ & $\bar{E}$& $\bar{C}^+_3$& $\bar{C}^-_3$ &$\bar{\sigma}_h$ & $\bar{S}^+_3$ & $\bar{S}^-_3$  \\ \hline
$A'$ &$K_1$                   & 1 & 1      &1        &1          & 1      & 1       &1         &1             &  1            &  1              &  1            &  1  \\
$^2 E'$ &$K_2$                & 1 &$\omega$&$\omega^*$ &1        &$\omega$&$\omega^*$ & 1      &$\omega$      &$\omega^*$     &1                &$\omega$       &$\omega^*$  \\
$^1 E'$ &$K_3$                & 1 &$\omega^*$&$\omega$ &1        &$\omega^*$&$\omega$ & 1      &$\omega^*$    &$\omega$       &1                &$\omega^*$     &$\omega$    \\
$A''$ &$K_4$                  & 1 & 1      &1        &$-1$       &$-1$    & $-1$      & 1      & 1            &1              &$-1$             &$-1$           & $-1$   \\
$^2 E''$ &$K_5$               & 1 &$\omega$&$\omega^*$ &$-1$     &$-\omega$&$-\omega^*$ & 1    &$\omega$      &$\omega^*$     &$-1$               &$-\omega$      &$-\omega^*$ \\
$^1 E''$ &$K_6$               & 1 &$\omega^*$&$\omega$ &$-1$     &$-\omega^*$&$-\omega$ & 1    &$\omega^*$    &$\omega$       &$-1$               &$-\omega^*$    &$-\omega$  \\
$^1\bar{E}_3$ &$K_7$          & 1 &$-\omega$&$-\omega^*$&$i$    &$-i\omega$&$i\omega^*$ & $-1$   &$\omega$      &$\omega^*$     &$-i$             &$i\omega$      &$-i\omega^*$ \\
$^2\bar{E}_3$ &$K_8$          & 1 &$-\omega^*$&$-\omega$&$-i$    &$i\omega^*$&$-i\omega$ & $-1$   &$\omega^*$   &$\omega$       &$i$              &$-i\omega^*$   &$i\omega$ \\
$^2\bar{E}_2$ &$K_9$          & 1 &$-\omega$&$-\omega^*$&$-i$    &$i\omega$  &$-i\omega^*$ & $-1$ &$\omega$     &$\omega^*$     &$i$              &$-i\omega$     &$i\omega^*$   \\
$^1\bar{E}_2$ &$K_{10}$       & 1 &$-\omega^*$&$-\omega$&$i$    &$-i\omega^*$ &$i\omega$  & $-1$ &$\omega^*$    &$\omega$       &$-i$             &$i\omega^*$    &$-i\omega$\\
$^1\bar{E}_1$ &$K_{11}$       & 1 &$-1$    &$-1$     &$i$        &$-i$    &$i$       & $-1$      &$1$           &$1$            &$-i$             &$i$            &$-i$       \\
$^2\bar{E}_1$ &$K_{12}$       & 1 &$-1$    &$-1$     &$-i$       &$i$     &$-i$       & $-1$      &$1$           &$1$            &$i$             &$-i$            &$i$      \\
\hline\hline
\end{tabular} }
\footnotetext{$\omega=\exp(2\pi i/3)$. }
\end{table}

\begin{table}[!htbp]
{ 
\caption{Character table at $T (\Gamma-K)$ axis, the $C_{1h}$ point double group \cite{Bradley_Cracknell72}. We replace $\Gamma$ with $T$  in the IR notation.}
\label{tab:CharactorTable_T}
\renewcommand{\arraystretch}{1.1}
\tabcolsep=0.1cm
\begin{tabular}{cc|cccc}
\hline \hline
\multicolumn{2}{c|}{$C_{1h}$} &$E$ &$\sigma_h$ & $\bar{E}$& $\bar{\sigma}_h$ \\ \hline
$A'$ &$T_1$                 & 1  & 1          &1           &1         \\
$A''$ &$T_2$                & 1 & $-1$         &1          &$-1$        \\
$^1\bar{E}$ &$T_3$          & 1 & $i$          &$-1$          &$-i$    \\
$^2\bar{E}$ &$T_4$          & 1 & $-i$         &$-1$          &$i$     \\
\hline\hline
\end{tabular} }
\end{table}

\begin{figure}
\includegraphics[width=8cm]{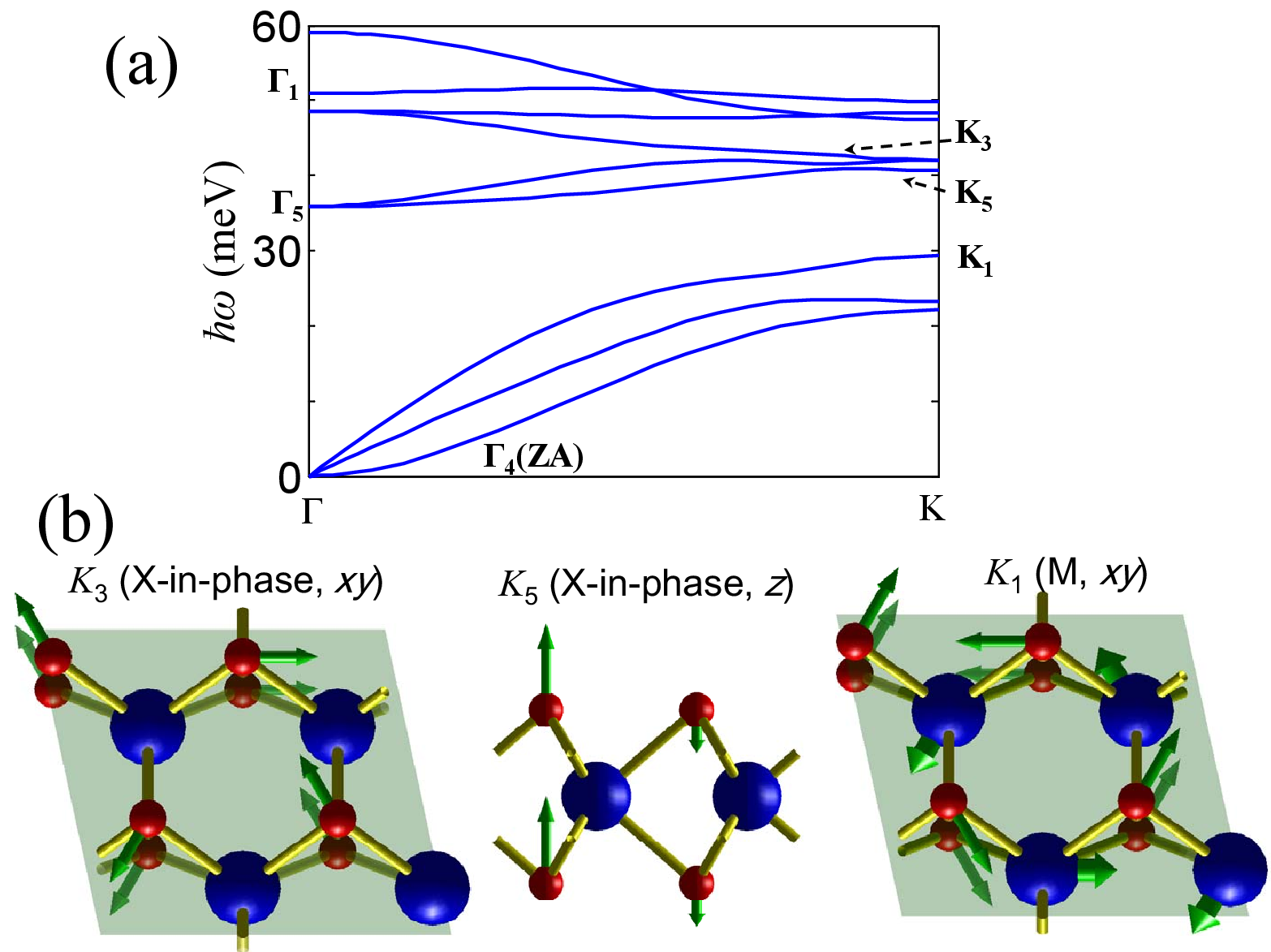}
\caption{(a) Phonon dispersion along $\Gamma$-$K$ line. Phonon modes that appear in Table~I of the main text are highlighted. (b) Atomic displacements of the zone-edge phonons. This information complements Fig.~2 of the main text.} \label{fig:phonon_band} \vspace{2mm}
\end{figure}

\begin{table*}[!htbp]
{ 
\caption{IR association of atomic orbitals in the upper valence band ($K_1$) and lower conduction bands ($K_{2-5}$) with respect to the $K$ point as defined in Fig.~1 of the main text. One needs to consider the lattice site factor in Eq.~(\ref{eq:tight_binding_state}). The $d$ orbitals from the transition-metal dominate these bands (so called metal $d$ band), while chalcogen $s$ and $p$ orbitals have smaller weights. We emphasize that they are `orbital-like' functions only in the sense of $C_{3h}$ operations and do not possess the spherical or inversion symmetry. For chalcogen X orbitals, the two layers can have phase 1 between them (polar, ``p'') or phase -1 between them (homopolar, ``h''). }
\label{tab:IR_orbitals}
\renewcommand{\arraystretch}{1.5}
\tabcolsep=0.5cm
\begin{tabular}{c|ccccc}
\hline \hline
IR &       $K_1$&$K_2$ & $K_3$& $K_4$& $K_5$ \\ \hline
M $d$-like orbitals & $d_{x^2-y^2}+id_{2xy}$ & $d_{x^2-y^2}-id_{2xy}$ & $d_{z^2}$  & $d_{xz}-id_{yz}$  & $d_{xz}+id_{yz}$        \\
X $s,p$-like orbitals & $p_x+ip_y$ (p)& $p_z$ (h), $s$ (p) & $p_x-ip_y$ (p) & $p_x+ip_y$ (h) & $p_z$ (p), $s$ (h) \\
\hline\hline
\end{tabular} }
\end{table*}
Phonon modes involved in Table~I and Fig.~2 of the main text are identified in Fig.~\ref{fig:phonon_band}. The phonon bands are generated using ABINIT DFT package, without taking into account the coupling between bending and stretching degrees of freedom. As a result, the dispersion of the ZA mode near $\mathbf{q}=\Gamma$ follows $\omega\propto q^\eta$ where $\eta\sim 2$.

$\\$$\\$

{\bf 2. Zone center valence band structure in Fig.~1, scattering matrix elements in Eq.~(1) and its spin-conserving counterpart, and spin-dependent scattering constants $a=\Xi^{so}_K$ in Eq.~(2).} With the IR identification of valley center states, band dispersion and states nearby are naturally obtained by the $\mathbf{k}\cdot\mathbf{p}$ method. The perturbation Hamiltonian for double group basis states is
\begin{eqnarray}
H_1=\frac{\hbar^2k^2}{2m_0}+ \frac{\hbar\mathbf{k}\cdot\bm{\pi}_\|}{m_0},
\end{eqnarray}
where $\bm{\pi}=\mathbf{p}+\hbar\bm\sigma\times\bm\nabla V/ 4m_0c^2$. $\pi_-(p_-)$ or $\pi_+(p_+)$ transforms as $K_2$ or $K_3$ IR at $\mathbf{K}$ point, as $\Gamma_6$ IR at $\Gamma$ point and as $T_1$ IR at $T$ point ($A_{\pm}$ denotes $A_x\pm iA_y$).  The spin splitting around $\Gamma_1$ of the valence band, shown in Fig.~1(b) of the main text, is a result of the three-fold rotational symmetry, and can be obtained from the method of invariants \cite{Bir_Pikus_Book},
\begin{eqnarray}
\Gamma^*_1\times \Gamma_1:\quad\sigma_z k_y(3k^2_y-k^2_x),\label{eq:energy_Gamma}
\end{eqnarray}
where $\sigma_z$ belongs to $\Gamma_2$ from Table~\ref{tab:CharactorTable_Gamma}, and $x$ direction connects two M atoms [Fig.~1(a) of the main text]. In order to derive spin-flip scattering constants in the following, we first find the spin mixing for conduction and valence band edges at $\mathbf{K}$,
\begin{subequations}\label{eq:spin_mix}
\begin{eqnarray}
|\mathbf{K}_c\Uparrow(\Downarrow)\rangle &=& |K_c \uparrow(\downarrow)\rangle + \frac{\Delta^{5(4)c}_{+(-)}}{E_{c}-E_{5(4)}}|K_{5(4)} \downarrow(\uparrow)\rangle,\label{eq:spin_mix_c}\qquad\\
|\mathbf{K}_v\Uparrow(\Downarrow)\rangle &=& |K_v \uparrow(\downarrow)\rangle + \frac{\Delta^{6(5)v}_{+(-)}}{E_{v}-E_{6(5)}}|K_{6(5)} \downarrow(\uparrow)\rangle,\label{eq:spin_mix_v}\qquad\\
\Delta^{nm}_{+(-)}&=&\frac{\hbar}{4m^2_0c^2}\langle K_n|(\bm\nabla V \times \mathbf{p})_{+(-)} |K_m\rangle,
\end{eqnarray}
\end{subequations}
where $\sigma_{x,y}$ belonging to $\Gamma_5$ is used in obtaining the selection rules. The dominant symmetry-allowed bands are shown in Table~\ref{tab:IR_orbitals}. Note that $K_6$ IR associates a farther band consisting M $p_z$ orbital and X $p_{x,y}$ orbital \cite{Mattheiss_PRB73}, and thus induces smaller spin mixing for valence band.

To compactly express the spin-conserving scattering rates  we employ the method of invariants \cite{Bir_Pikus_Book}. The transition amplitudes for spin-conserving scattering with in-plane long-wavelength acoustic phonons in the $\Gamma$, $K$ or $T$ valleys are
\begin{eqnarray}
\mathcal{T}_{\lambda,\mathbf{q}}^{sc} \!=\! \sqrt{\frac{k_BT}{2\varrho A v_{\lambda}^2q^2}}\! \left\{\!\!\begin{array}{l} \,\Xi_{\Gamma_1}\bm\xi_{\lambda}\cdot \mathbf{q}, \vspace{1.5mm} \\  \,\Xi_{K_{1(3)}}^{d} \bm\xi_{\lambda}\cdot \mathbf{q}+ \Xi_{K_{1(3)}}^{o}(\bm\xi_{\lambda} \times \mathbf{q})_z,  \vspace{1.5mm} \\  \,\Xi_{T_1}^{\alpha\beta}\xi_{\lambda,\alpha}q_{\beta}, \end{array}\right.\quad \label{eq:Tsc}
\end{eqnarray}
where $\varrho$, $A$ and $v_{\lambda}$ are the areal mass density, area, and mode-dependent sound velocity, respectively. $\mathbf{q}=\mathbf{k}_f-\mathbf{k}_i$ is the small wavevector difference between the initial and final states. Summation over tensor components is implied in the $T$-valley. The dilation ($\Xi^{d}_{K_1}$) and shear ($\Xi^{o}_{K_1}$) deformation potentials of the conduction band at $K$ point have been numerically fit to be 2.8 eV and 1.6 eV for MoS$_2$ \cite{Kaasbjerg_PRB12}. Other constants are of the same order of magnitude, by considering the majority of orbitals involved in these bands. The clear physical meaning of these scattering constants considerably simplifies their measurement numerically or experimentally \cite{Song_PRB12}.

The forms of the spin-flip matrix elements core  $g(\mathbf{k}_i, \mathbf{k}_f)$ in Eq.~(1) of the main text are also obtained by the method of invariants. In $T$ valley, the zeroth-order term has been identified by the direct selection rule in Table~I of the main text.  In conduction (valence) $\mathbf{K}$ valley we have
\begin{eqnarray}
K^*_{9(8)}\times K_{11(7)}=K_6&:& \quad\xi_z q_+\;\; \textrm{at } \mathbf{K}_{c(v)}.\,
\end{eqnarray}
This selection rule also governs the out-of-plane shear strain ($\epsilon_+$ and $\xi_z q_+$ transform the same). Together with the spin-conserving scattering that is enabled by two-flexural-phonon process (transforms as $\epsilon_+^2$) \cite{Mariani_PRL08}, they set the leading order terms in the spin-dependent strain Hamiltonian [Eq.~(3) of the main text]. The magnitudes of their associated \textit{spin-dependent deformation potential constants} $a=\Xi^{so}_K$ are crucial in determining the fast spin relaxation. These constants can be estimated analytically by identifying the majority orbitals for the involved bands [see Eqs.~(\ref{eq:spin_mix}) and Table~\ref{tab:IR_orbitals}].  We find large contributions from the Elliott part clearly, and have
\begin{subequations}
\begin{eqnarray}
\Xi^{so}_{K_c} &\approx& -\frac{i}{2}N\left( \frac{\Delta^{5c}_+}{E_c-E_5}\left\langle K_c \left|r_-\frac{\partial V^{\rm{at}}_+}{\partial z} \right| K_5 \right\rangle \right.\nonumber \label{eq:spin-dep-deformation-c}\\
&&\left. \;\;\qquad + \frac{\Delta^{c4}_+}{E_c-E_4}\left\langle K_4 \left|r_-\frac{\partial V^{\rm{at}}_+}{\partial z} \right| K_c \right\rangle \right),\quad\\
\Xi^{so}_{K_v} &\approx& -\frac{i}{2}N \frac{\Delta^{v5}_+}{E_v-E_5}\left\langle K_5 \left|r_-\frac{\partial V^{\rm{at}}_+}{\partial z} \right| K_v \right\rangle ,
\end{eqnarray}
\end{subequations}
where $V_+$ is the sum of atomic potentials in a primitive cell. By taking the dominant $K_{4,5}$ orbitals in Table~\ref{tab:IR_orbitals}, we can approximate the interband spin-orbit coupling by the atomic spin splittings ($|\Delta_+|\!\sim\! 0.2$ eV), and consequently $\Xi^{so}_{K_{c(v)}}\!\!\sim\! 0.2$ eV. Note that \textit{the large $\Xi^{so}_{K_{c}}$ comes directly from the difference in the two energy denominators in Eq.~(\ref{eq:spin-dep-deformation-c}) for conduction band, which reflect the strength of crystal potential that is odd under space inversion.} This order of magnitude is confirmed with DFT calculation.  With conduction band spin splitting $\Delta_K$\,$\gg$\,$|a\epsilon_+|$ and $a$\,$\sim$ \,$b$\,$\gg$\,$\Delta_K$, we safely simplify the strained splitting to $\sqrt{\Delta^2_K + |2a\epsilon_+|^2}$ and extract $a$.

Finally, at $\Gamma$ valence band the quadratic-in-wavevector $g(\mathbf{k}_i, \mathbf{k}_f)$ of Eq.~(1) has the form
\begin{eqnarray}
\Gamma^*_1\times \Gamma_1\!=\!\Gamma_1:  \xi_z\sigma_+\![\bar{k}_x q_x\! -\!\bar{k}_y q_y\! +\!i (\bar{k}_x q_y\!+\!\bar{k}_y q_x)],\quad
\end{eqnarray}
where $\bar{\mathbf{k}}$ is the average of initial and final wavevector. The wavevector dependence is $\bar{k}q$ in terms of magnitude. Similar to $\Gamma$ valley in zinc-blende crystals, when $\bar{\mathbf{k}}=0$ the spin flip vanishes due to time reversal symmetry; when $\bar{\mathbf{k}} \neq 0$ but $\mathbf{q}=0$, the spin-flip matrix element is negligibly higher-order in wavevector \cite{Yafet_SSP63}.

$\\$$\\$

{\bf 3. Excitons and optical transition rate.} A direct (indirect) exciton band is associated with a pair of $c$ and $v$ bands around the same (opposite) $K$ point, with energy $E_{\mathcal{X}}(\bm{\mathcal{K}}) = E_g - E_{b,1s}+\hbar^2 |\bm{\mathcal{K}} -\bm{\mathcal{K}_i}|^2/2 M$ in effective mass approximation \cite{Dresselhaus_JPCS56}. $E_g$, $E_{b,1s}$ and $M=m_e+m_h$ are the associated $c{\mbox -}v$ band gap, binding energy and effective mass respectively.  $E_{b,1s}$ is found to be the order of eV, so only $1s$ optically accessible bound states are concerned here. $\bm{\mathcal{K}}_i=0,\pm\mathbf{K}$ is the respective exciton valley center. The exciton state $\Psi_{\bm{\mathcal{K}}}$ is the product of single particle states $\psi_{c}$, $\psi_{v}$ from proper electron and hole $K$ valleys ($\bm{\mathcal{K}}= \mathbf{k}_e +\mathbf{k}_h$, with $\mathbf{k}_h$ being \textit{negative} to the wavevector of valence band state $\psi_{v,-\mathbf{k}_h}$  in the single electron band diagram), modulated by the hydrogen-like envelope function \cite{Dresselhaus_JPCS56,Eliott_PR57},
\begin{subequations}
\begin{eqnarray}
\label{eq:direct_exciton}
\Psi^\pm_{\bm{\mathcal{K}}=0}(\mathbf{r}_e,\mathbf{r}_h)& \approx & \frac{1}{\sqrt{S}}\Phi_{1s}(\mathbf{r})
\psi_{c,\pm\mathbf{K}}(\mathbf{r}_e) \psi_{v,\pm\mathbf{K}}(\mathbf{r}_h),\qquad\\
\label{eq:indirect_exciton}
\Psi_{\bm{\mathcal{K}}=\pm\mathbf{K}}(\mathbf{r}_e,\mathbf{r}_h)& \approx & \frac{1}{\sqrt{S}}\Phi_{1s}(\mathbf{r})
\psi_{c,\mp\mathbf{K}}(\mathbf{r}_e) \psi_{v,\pm\mathbf{K}}(\mathbf{r}_h),\qquad
\end{eqnarray}
\end{subequations}
where $S$ is the area of the 2D crystal, $\Phi_{1s}$ is the 2D $1s$ hydrogenic-like wavefunction [with $\mu=(\frac{1}{m_e}+\frac{1}{m_h})^{-1}$ and $\epsilon\epsilon_0$], and $\mathbf{r}=\mathbf{r}_e-\mathbf{r}_h$. $\Psi^\pm_{\bm{\mathcal{K}}=0}$ are doubly-degenerate direct excitons excited by $\sigma=\pm$ circularly polarized light, belonging to $\Gamma_6$ IR,
\begin{subequations}
\begin{eqnarray}
\Gamma_6 & =&\{K_{11}\times K^*_7, K_{12}\times K^*_8\}, \quad \!\!\textrm{direct }\mathcal{X}_A \nonumber\\
&=&\{K_9\times K^*_8, K_{10}\times K^*_7\}, \quad \textrm{direct }\mathcal{X}_B
\end{eqnarray}
from IRs of their constituent states [see Fig.~1(b)]. They are bright excitons and effective in optical transitions. Also ``bright'' are indirect excitons at $\pm\mathbf{K}$ valleys connected with direct excitons by fast spin-conserving scattering, belonging to $K_{2,3}$ IR (see Fig.~3),
\begin{eqnarray}
K_2 & =&  K_{12}\times K^*_8 =K_{10}\times K^*_7,  \nonumber\\
K_3  &=&  K_{11}\times K^*_7 =K_{9}\times K^*_8.
\end{eqnarray}
\end{subequations}

\begin{figure}
\includegraphics[width=8cm]{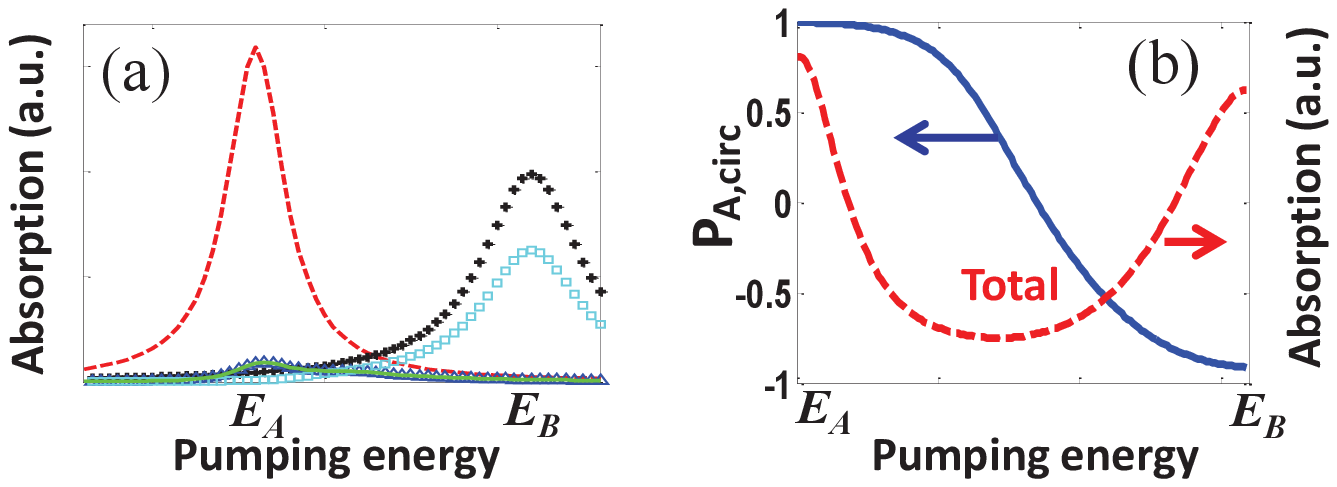}
\caption{(Color online) (a) Absorption rates as functions of $\sigma_+$ photon pumping energy for direct $\mathcal{X}_{A}$ (red dash), direct $\mathcal{X}_{B}$ (black plus), indirect $\mathcal{X}_{A}$ in $\Gamma$ valley (green solid), indirect $\mathcal{X}_{A}$ in $\mathbf{K}$ valley (blue wedge) and indirect $\mathcal{X}_{A}$ in $-\mathbf{K}$ valley (cyan square). (b) Total absorption rate and PL helicity as functions of pumping energy.} \label{fig:exciton}
\end{figure}

As explained in the main text, the fast spin-conserving exciton scattering is mainly responsible for the photoluminescence (PL) circularity and can be shown by straightforward rate equations. In the case of MoS$_2$, the experimental results generally show a wide ($\sim$50 meV) absorption peak \cite{Mak_NatMat13}, and $\mathcal{X}_B$ absorption tail needs to be considered. Substrate inhomogeneity, impurity and phonon scattering contribute to this broadening. We focus on the low-temperature regime where the lineshape is of Lorentzian shape. Only phonon emission needs to be considered and indirect photon absorption to $\mathcal{X}_B$ is to be neglected (for excitation energy $\hbar \omega_\nu<E_{\mathcal{X}_B}$). The absorption rate to direct $\mathcal{X}_i$ ($i=A,B$) by circularly polarized ($\sigma=\pm$) light is
\begin{subequations}
\begin{eqnarray}\label{eq:direct_absorb}
R^{\rm dir}_i(\omega_\nu) &\approx & \frac{2\pi}{\hbar} |\langle \Psi^\sigma_{0,i} | \mathcal{H}^{\sigma}_{\rm{eR}} |0\rangle|^2 \mathcal{L}[\hbar\omega_\nu;E_{\mathcal{X}_i}(0)/, \hbar\Gamma_i/2]\nonumber\\
&\approx & C_i \mathcal{L}[\hbar\omega_\nu;E_{\mathcal{X}_i}(0), \hbar\Gamma_i/2],
\end{eqnarray}
where $|0\rangle$ is the ground state, $\Gamma_i$ is the linewidth, and  $C_i=\hbar[e\mathcal{E}|P_{cv_i}|/m_0 E_{\mathcal{X}_i}(0)a_{Bi}]^2$, with electric field $\mathcal{E}$, interband momentum matrix element $P_{cv_i}$ and Bohr radius $a_{Bi}$ associated with $\mathcal{X}_i$. The absorption rate to indirect $\mathcal{X}_i$ in $\pm\mathbf{K}$  valley via direct $\mathcal{X}_j$ virtual state, by circularly polarized $\sigma$ light is
\begin{eqnarray}\label{eq:indirect_absorb}
R^{j\!\rightarrow\! i}( \omega_\nu) &\approx & \frac{2\pi}{\hbar} \sum_{\bm{\mathcal{K}}} \left| \frac{\langle \Psi_{\bm{\mathcal{K}},i} |\mathcal{H}_{\rm ep} |\Psi^\sigma_{0,j}\rangle \langle \Psi^\sigma_{0,j}| \mathcal{H}^{\sigma}_{\rm{eR}} |0\rangle}{E_{\mathcal{X}_j}(0)-\hbar \omega_\nu + i \hbar \Gamma_j/2} \right|^2  \quad\nonumber\\
&& \widetilde{\Theta}_i(\omega_\nu-\omega_{\mathbf{K}}- E_{\mathcal{X}_i}(\mathbf{K})/\hbar)  \nonumber\\
& \approx &   C_j  \frac{M_i} {\pi\hbar\rho\omega_{\mathbf{K}}} \left| \frac{V^{ij}_{\rm ep}}{{E_{\mathcal{X}_j}(0)-\hbar \omega_\nu} + i \hbar \Gamma_j/2}\right|^2 \nonumber\\
&&(n_{\mathbf{K}}+1)\widetilde{\Theta}_i(\omega_\nu-\omega_{\mathbf{K}}- E_{\mathcal{X}_i}(\mathbf{K})/\hbar),\qquad
\end{eqnarray}
where $|V^{ij}_{\rm ep}|\! \approx\! N |\langle \psi_{n, \!-\mathbf{K}}|\!\sum_{\alpha} \bm\xi_{\alpha}\! \cdot\! \bm\nabla V_{\alpha}(\mathbf{r}\!-\!\bm{\tau}_\alpha) e^{i \mathbf{K}\cdot \bm{\tau}_\alpha} |\psi_{n, \mathbf{K}}\rangle|$, with $\alpha$ indexing the unit cell atom,  $n=c(v)$ if $i=(\neq)j$ and $\psi$ not including spin; $\widetilde{\Theta}_i(\omega)$ denotes a broadened step function $\arctan(2\omega/\Gamma_i)/\pi+1/2$; $M_i$ is the effective mass of $i$th exciton, and $\omega_{\mathbf{K}}$ and $n_{\mathbf{K}}$ are the phonon frequency and population at wavevector $\mathbf{K}$ respectively. $E_{\mathcal{X}_i}(\mathbf{K})\approx E_{\mathcal{X}_i}(0)$. Similarly the indirect absorption rate to $\mathcal{X}_i$ in central valley ($\Gamma_1$ blue arrow in Fig.~3) is
\begin{eqnarray}\label{eq:indirect_absorb_intra}
R^{\Gamma}_{i}(\omega_\nu)
& \approx &   C_i  \frac{M_i} {\pi\hbar\rho\omega_{0}} \left| \frac{V^{\Gamma}_{\rm ep}}{{E_{{\rm x}i}(0)-\hbar \omega_\nu} + i \hbar \Gamma_i/2}\right|^2 ,\qquad
\end{eqnarray}
\end{subequations}
where $|V^{\Gamma}_{\rm ep}|\approx N|\frac{1}{2}\sum_n\langle \psi_{n, \mathbf{K}}|\sum_{\alpha} \bm\xi_{\alpha} \cdot \bm\nabla V_{\alpha}(\mathbf{r}-\bm{\tau}_\alpha)  |\psi_{n, \mathbf{K}}\rangle|$ with $n=c,v$.  To calculate the individual absorption spectra in MoS$_2$, we specify $R^{\rm dir}_{A,B}$ in Eq.~(\ref{eq:direct_absorb}),  $R^{j\!\rightarrow\! A}$ in Eq.~(\ref{eq:indirect_absorb}) with $K_{3(1)}$ phonon for $j=A(B)$, and $R^{\Gamma}_{A}$ in Eq.~(\ref{eq:indirect_absorb_intra}) with $\Gamma_1$ (out-of-plane homopolar) phonon. The results are plotted in Fig.~\ref{fig:exciton}(a) using ballpark values $|V^{ij}_{\rm{ep}}|,|V^{\Gamma}_{\rm{ep}}|\sim$ 2 eV$(2\pi/a)$, and $\hbar \omega_{K1}=20$ meV, $\hbar \omega_{K3}=50$ meV and $\hbar \omega_{\Gamma_1}=50$ meV [Fig.~\ref{fig:phonon_band}(a)]. The identical shapes from indirect and direct absorptions are caused by the constant 2D density of states, and it obstructs the experimental resolution of the indirect contribution. The total absorption in Fig.~\ref{fig:exciton}(b) matches experimental results for absorption of neutral excitons \cite{Mak_NatMat13}.

Next we calculate the PL around $\mathcal{X}_A$ peak. After the excitons are generated, they undergo intravalley and intervalley scattering before emission. As pointed out in the main text, excitons via direct $\mathcal{X}_B$ state end up in direct $\mathcal{X}_A$ with opposite magnetic number $-\sigma$. \textit{At this range, the indirect emission is negligible as there is no advantage of availability in the final states}.  Most of indirect excitons go back to ground state by non-radiative emission. The rate equation for steady state density $N$ under $\sigma$ excitation light are
\begin{subequations}
\begin{eqnarray}
\frac{d N^{\rm dir}_{A,\sigma} }{dt} &=& R^{\rm dir}_A (\omega_\nu)+ \frac{R^\Gamma_A(\omega_\nu)}{1+ \widetilde{\Theta}_A(\omega_\nu-\omega_1)/\gamma} \nonumber\\ &&\!\! + N^{A\!\rightarrow \!A} R^{{\rm inter}}_{c}
\widetilde{\Theta}_A(\omega_\nu-\omega_2)\!\!- \!\!N^{\rm dir}_{A,\sigma} \widetilde{R}^{\rm dir}_A,\qquad \\
\frac{d N^{\rm dir}_{A,-\sigma}}{dt} &=& \!\!- \!N^{\rm dir}_{A,-\sigma} \widetilde{R}^{\rm dir}_A \!\!+ \!\!N^{B\!\rightarrow\! A} R^{\rm inter}_{c} \widetilde{\Theta}_A(\omega_\nu-\omega_3),\;\;\qquad\\
\frac{d N^{A\!\rightarrow \!A}}{dt} &=& \!R^{A\!\rightarrow\! A} (\omega_\nu)\nonumber\\
 &&- N^{A\!\rightarrow \!A}[ R^{\rm intra}+ R^{\rm inter}_{c} \widetilde{\Theta}_A(\omega_\nu-\omega_2)],\quad\qquad\\
\frac{d N^{B\!\rightarrow\! A}}{dt} &=& R^{B\!\rightarrow\! A} (\omega_\nu)\!+ \!N^{\rm dir}_{B,\sigma} R^{\rm inter}_{v} \!\!- \!\!N^{B\!\rightarrow \!A}[ R^{\rm intra} \quad\qquad\nonumber\\  &&+R^{\rm inter}_{c} \widetilde{\Theta}_A(\omega_\nu-\omega_3)],\qquad\\
\frac{d N^{\rm dir}_{B,\sigma} }{dt} &=& R^{\rm dir}_B (\omega_\nu) - N^{\rm dir}_{B,\sigma} \widetilde{R}^{\rm dir}_B + N^{\rm dir}_{B,\sigma} R^{\rm inter}_{v},\quad\qquad\; \end{eqnarray} \end{subequations}
where $N^{\rm dir}_{A,B}$ denotes direct $\mathcal{X}_{A,B}$ density and  $N^{i\!\rightarrow \!j}$ the indirect $\mathcal{X}_{j}$ density via direct $\mathcal{X}_{i}$,  $\widetilde{R}$ is the photon emission rate, $R^{\rm inter}_{c,v}$ and $R^{\rm intra}$ are the fast intervalley and intravalley scattering rates, $\gamma = R^{\rm intra} /R^{\rm inter}_{c}$ is on the order of 1,
$\omega_1 =E_{{\rm x}A}(0)/\hbar+ \omega_{\Gamma_1}+\omega_{K_3}$, $\omega_2 = E_{{\rm x}A}(0)/\hbar+ 2\omega_{K_3}$, and $\omega_3 = E_{{\rm x}A}(0)/\hbar+\omega_{K_1}+\omega_{K_3}$. Scattering rates as well as photon emission rates drop out when the above equations are solved to obtain the ratio $N^{\rm dir}_{A,\sigma}/N^{\rm dir}_{A,-\sigma}$, and the average PL helicity at the $\mathcal{X}_A$ peak follows
\begin{subequations}
\begin{eqnarray}
\bar{P}_{\rm circ}(\omega_A) &=& \frac{\bar{P}_{\sigma}(\omega_A) - \bar{P}_{-\sigma}( \omega_A)}{\bar{P}_{\sigma} ( \omega_A) + \bar{P}_{-\sigma} ( \omega_A)}\approx  \frac{N^{\rm dir}_{\sigma} - N^{\rm dir}_{-\sigma}}{N^{\rm dir}_{\sigma} + N^{\rm dir}_{-\sigma}},\;\qquad
\end{eqnarray}
with
\begin{eqnarray}
\frac{N^{\rm dir}_{\sigma}}{N^{\rm dir}_{-\sigma}} &\approx &
\frac{R^{\rm dir}_A (\omega_\nu)+ R^{A\!\rightarrow\! A}(\omega_\nu)/ d_2
+ R^{\Gamma}_A(\omega_\nu)/d_1}{[R^{B\!\rightarrow \!A}(\omega_\nu) + R^{\rm dir}_B (\omega_\nu) ]/ d_3 },\qquad\;
\end{eqnarray}
\end{subequations}
where $d_1 =1+1/[\gamma\widetilde{\Theta}_A(\omega_\nu-\omega_1) ]$ and $d_{2(3)}= 1+\gamma/\widetilde{\Theta}_A(\omega_\nu-\omega_{2(3)})$.    The result is plotted in Fig.~\ref{fig:exciton}(b) as a function of excitation energy $\hbar \omega_\nu$. The PL circularity decreases with increasing $\hbar \omega_\nu$  since more excitons go via upper $\Gamma_6$ band. The negative result near $\hbar \omega_\nu=E_B$ is our prediction in the neutral exciton case. In experiments, on the other hand, the polarization does not change sign but drops to zero. This case is explained from the involvement of trions (charges exciton), where half of the trions with energies above the $B$ level are from the tail of the $A$-exciton band \cite{Mak_NatMat13}.